\documentclass{vgtc}                          

\ifpdf
  \pdfoutput=1\relax                   
  \usepackage{graphicx}                
  \DeclareGraphicsExtensions{.pdf,.png,.jpg,.jpeg} 
\else
  \usepackage{graphicx}                
  \DeclareGraphicsExtensions{.eps}     
\fi%

\graphicspath{{figures/}{pictures/}{images/}{./}} 

\usepackage{microtype}                 
\PassOptionsToPackage{warn}{textcomp}  
\usepackage{textcomp}                  
\usepackage{mathptmx}                  
\usepackage{times}                     
\usepackage{cite}                      
\usepackage{tabu}                      
\usepackage{booktabs}                  
\usepackage{xcolor}
\usepackage{amsmath}
\usepackage{dsfont}
\usepackage{amsfonts}

\onlineid{0}

\vgtccategory{Research}

\vgtcinsertpkg

\title{SAX Navigator: Time Series Exploration through Hierarchical Clustering}

\author{Nicholas Ruta\thanks{e-mail: nruta@g.harvard.edu}\\ %
        \scriptsize Harvard University %
\and Naoko Sawada\thanks{e-mail: nsawada@fj.ics.keio.ac.jp}\\ %
     \parbox{1.4in}{\scriptsize \centering Harvard University \\ Keio University}
\and Katy McKeough\thanks{e-mail: kathrynmckeough@g.harvard.edu}\\ %
     \scriptsize Harvard University %
\and Michael Behrisch\thanks{e-mail: behrisch@g.harvard.edu}\\  %
     \scriptsize Harvard University %
\and Johanna Beyer\thanks{e-mail: jbeyer@g.harvard.edu}\\  %
     \scriptsize Harvard University
}

\teaser{
  \centering
  \includegraphics[width=\linewidth]{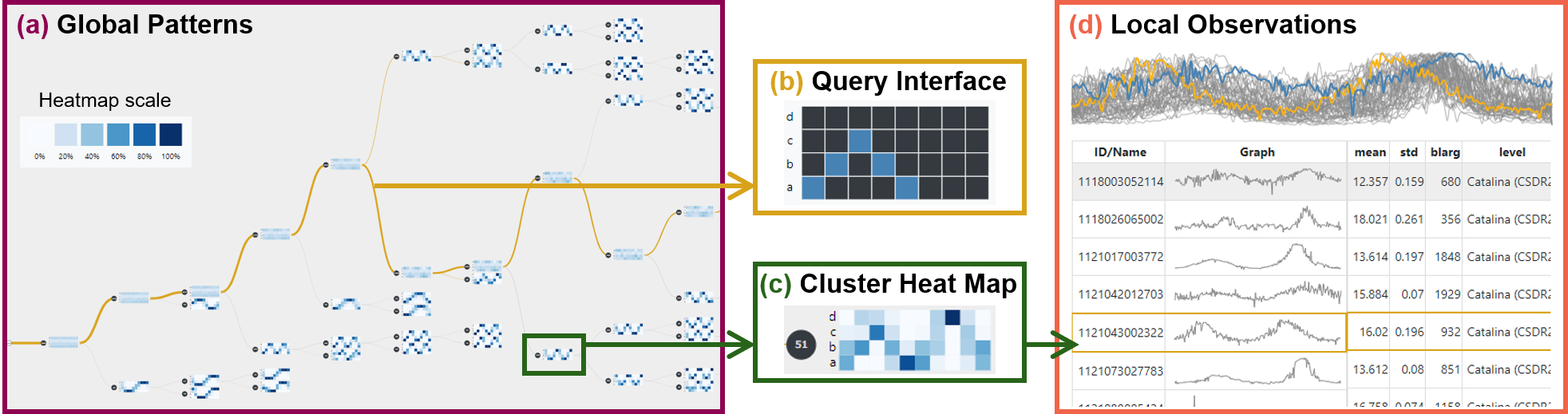}
  \caption{SAX Navigator shows the hierarchical clustering result for 2{,}000 astronomical observations (i.e., time series). Tree diagram (a) showing the global patterns derived from the hierarchical clustering of all time series. Tree branches are highlighted based on the user-specified pattern expressed in the visual query interface (b). Each tree node features a cluster heat map (c) representing the general shape of all time series in the cluster. A details-on-demand display (d) shows local observations of a single cluster.}
	\label{fig:teaser}
}

\abstract{
Comparing many long time series is challenging to do by hand. Clustering time series enables data analysts to discover relevance between and anomalies among multiple time series. However, even after reasonable clustering, analysts have to scrutinize correlations between clusters or similarities within a cluster. 
We developed SAX Navigator, an interactive visualization tool, that allows users to hierarchically explore global patterns as well as individual observations across large collections of time series data. 
Our visualization provides a unique way to navigate time series that involves a ``vocabulary of patterns'' developed by using a dimensionality reduction technique, \emph{Symbolic Aggregate approXimation} (SAX). With SAX, the time series data clusters efficiently and is quicker to query at scale. 
We demonstrate the ability of SAX Navigator to analyze patterns in large time series data based on three case studies for an astronomy data set.
We verify the usability of our system through a think-aloud study with an astronomy domain scientist. 
} 

\CCScatlist{
  \CCScatTwelve{Human-centered computing}{Visu\-al\-iza\-tion}{Visu\-al\-iza\-tion techniques}{Treemaps};
  \CCScatTwelve{Human-centered computing}{Visu\-al\-iza\-tion}{Visualization design and evaluation methods}{}
}

\begin{document}


\firstsection{Introduction}

\maketitle

Time series analysis is one of the most common analyses in a variety of domains. 
%
Time series analyses from a collection of observations over different timelines are much richer, yet more complex than those from a single observation.
%
An astronomer, for example, may be interested in searching for reoccurring patterns or anomalies
in the brightness over time across hundreds of thousands of different measurements of celestial bodies.
In sports, a coach might want to compare the career trajectories of different athletes.

Oftentimes, correlations and relationships in data are not internalized and understood through raw data, but rather through the apparent (visual) patterns they express. Therefore, users can benefit immensely from pattern-based navigation for the exploration of large collections of time series data.
Many existing techniques that explore patterns algorithmically, such as autoregressive models or Fourier transforms~\cite{LinBook2012}, focus on patterns at a global scale. 
Algorithms that analyze time series data globally are powerful in finding common patterns, but often do not account for \emph{why} these patterns are important. 
 On the other hand, visual analytics methods, such as the one presented by Correll and Gleicher \cite{Correll2016}, tend to solely view time series data at a local scale. 
 Only at a local level, we can answer questions about \emph{what} makes a particular data point an outlier or \emph{how} one cluster of timelines compares to another.
 Consequently, there is a need for efficient hybrid time series exploration techniques that extract patterns at a global scale, while still allowing for local exploration of the data.
Few techniques are capable of capturing global patterns and navigating local connections, such as Clustrophile 2~\cite{Cavallo2019}. Unfortunately, its approach cannot scale to dozens or hundreds of clusters due to perceptual scalability.

To solve the above challenges, we follow Symbolic Aggregate approXimation (SAX), which was presented by Lin et al.~\cite{Lin2007} to describe and simplify time series as a series of words from an automatically derived vocabulary (see~\autoref{fig:sax}). 
By means of SAX, we can retrieve patterns at a global scale via clustering using a distance function that is invariant for translation and scale. 

In this paper, we present SAX Navigator, a scalable visualization technique for analyzing large collections of time series data based on the hierarchical composition of its visual pattern space.
During the development of our tool, we focused on leveraging well-developed and evaluated techniques for pattern detection and extraction.
Therefore, the algorithmic portion of our tool utilizes hierarchical clustering, which has been proven to be a robust method. We utilize the dimensionality reduction method SAX, which makes for efficient querying and matching within our data by applying techniques already developed for regular expressions. 
Users can explore global patterns in the SAX Navigator tree view (see \autoref{fig:teaser} (a)), sketch a specific query (\autoref{fig:teaser} (b)), compare general trends and mean shapes of individual clusters in the cluster's heat map (see \autoref{fig:teaser} (c)), and look at the detailed cluster membership (see \autoref{fig:teaser} (d)).
\section{Related Work}\label{Sec: Related work}
We first review methods to visually query time series data to extract patterns and then describe visualization methods for clustered data. 

\noindent
\textbf{Query Definition for Time Series Analysis.}
Query-by-example and query-by-sketch interfaces are powerful approaches to querying data intuitively. 
%
Query-by-example techniques~\cite{DBLP:journals/ivs/HochheiserS04,DBLP:journals/tkde/WangDS12}
aim to find similar data points (e.g., time slices) to a user-specified example. However, they do not address how to find the initial interesting time slice from a large collection of time series data.
Query-by-sketch techniques do not have this restriction, as users can directly draw the shape they are interested in. However, query-by-sketch techniques have to deal with the user-introduced uncertainties of sketches~\cite{Correll2016}.


TimeSearcher is a visual exploration tool for time series data~\cite{DBLP:journals/ivs/HochheiserS04}, which is extended to a query-by-example interface, named SearchBox~\cite{Buono2008}. 
SOMFlow \cite{Sacha2018} presents techniques for time series clustering based on query-by-example, grouping selections based on their relative neighborhoods and by filtering and splitting using metadata-based attribute values. 
QuerySketch~\cite{Wattenberg2001} is a tool for database queries where users can directly sketch the shape of a pattern which automatically extracts matching time slices.
Correll and Gleicher~\cite{Correll2016} 
defined a vocabulary of invariants for queries by sketch to deal with uncertainties of sketches.

While these query-by-sketch systems allow users to draw and query time variation in an arbitrary shape, SAX Navigator provides users with building blocks that can be pieced together to create query examples based on \emph{observed} patterns in the data. It provides a comprehensive exploration of time series collections using query-by-example for specific observations of interest and query-by-sketch to collect results based on a general trend.

\noindent
\textbf{Visual Cluster Analysis.}
%
We will now survey a selection of tools that inspired the design of SAX Navigator. Seo and Shneiderman~\cite{Seo2002}, for example, presented the Hierarchical Clustering Explorer (HCE), a dendrogram-based interactive visual exploration tool for hierarchical clustering. It allows users to filter clusters according to similarities and to compare clusters. 
NodeTrix~\cite{Henry2007} solves the complexity of node-link diagrams of large networks by aggregating nodes into clusters and displaying dense clusters as matrices within the overall node-link diagram. 
CyteGuide~\cite{Hollt2018} enables users to explore the hierarchical representation of the data by viewing both the current status of exploration and the unexplored parts based on sunburst diagrams. 
Clustervision~\cite{Kwon2018} is a VA tool to help users find a proper clustering method from various techniques and parameters. 
Zeckzer et al.~\cite{Zeckzer2018} proposed tiled binned clustering and visualize the results in 3D scatterplots. The method conducts clustering after assigning data points to bins, as we do in SAX Navigator.

All these VA approaches efficiently show clusters and allow users to explore the cluster space. 
However, it is still often difficult to gain a comprehensive overview of the individual data samples contained in a cluster.
Similar to NodeTrix, we aim to reduce visual complexity by showing general patterns within clusters rather than individual observations.
We, therefore, incorporate a heat map-based cluster aggregation view into SAX Navigator.


%
\section{A Vocabulary of Patterns (SAX)}

The basis of our vocabulary of patterns revolves around a time series dimensionality reduction technique called Symbolic Aggregate approXimation (SAX) \cite{Lin2007}. SAX allows users to control the resolution of their analysis, 
but also to apply established and well-understood natural language processing techniques, such as text similarity and retrieval through regular expressions or topic analysis.

\begin{figure}[t]
    \centering
    \includegraphics[width=.99\columnwidth]{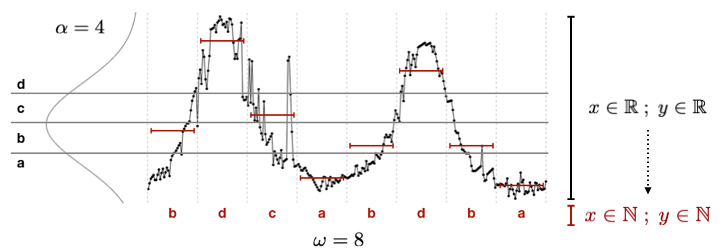}
    \caption{\footnotesize{Transforming a timeline into SAX representation with $\alpha=4$ letters and a word length of $\omega=8$. The dimension of the value ($x$) and time ($y$) are reduced from $\mathbb{R}$ to $\mathbb{N}$.}}
    \vspace{-4mm}
    \label{fig:sax}
\end{figure}

\subsection{SAX}

To prepare our data, we first center and scale it (i.e., we subtract out the mean and divide by the standard deviation). However, depending on the data characteristics, different pre-processing techniques might be used.
\autoref{fig:sax} depicts an example of translating/converting a time series into the SAX representation. 
Conceptually, SAX quantizes a continuous time series into discrete bins (along both, the time and amplitude axis) and assigns a letter representation to each quantized bin.
The first step to convert a time series into the SAX representation is to define the number of letters $\alpha$ and maximum word length (subsequent bin size) $\omega$ to apply to the data set. Both should be chosen to be the smallest possible values while allowing for good clustering and not smoothing away the details. To determine the distribution of the letters, SAX pools the values of all time series together and fits a normal distribution. Then it creates $\alpha$ partitions of equal probability and assigns the lowest to the letter ``a", the second lowest to ``b" and so on to create the set of letters in our vocabulary.
Some observations may not be $\omega$ letters long since not all time series have to be of the same length. 
Binning, a form of smoothing that removes noise from the data, improves the ability of the clustering algorithm to find similar groups of time series. For each bin, we average the values and determine its letter range.

The result is that each observation is a set of $\alpha$ letters of maximum length. We cluster the resulting words with the goal to find groups of time series with similar words within our vocabulary. Lastly, dimensionality reduction promotes scalability by decreasing the complexity of the time series from the space of $\mathbb{R}^2$ to $\mathbb{N}^2$.




\subsection{Clustering}
We use agglomerative hierarchical clustering with complete linkage for clustering time series into similar groups, since it heuristically provides better cluster separation than single or average linkage.  
The used distance metric is a variation of the MINDIST function, described in Lin et al.~\cite{Lin2007}, which achieves exact matching even though SAX words may contain empty values in our data set. It performs better than Euclidean distance in terms of recovering cluster assignments.
The distance between two time series observations as SAX representations $(S^{(1)}, S^{(2)})$ is defined as follows:

$$D(S^{(1)}, S^{(2)}) = 1- \frac{1}{\omega}\sum_{i=1}^\omega d(S^{(1)}_i, S^{(2)}_i)$$
\[d(S^{(1)}_i, S^{(2)}_i) =  \begin{cases} 
      1 & S^{(1)}_i = S^{(2)}_i  \\
      0 & S^{(1)}_i \mbox{is NaN} \; \cup \; S^{(2)}_i \mbox{is NaN} \\
      -1 & S^{(1)}_i \neq S^{(2)}_i 
  \end{cases}
\]
\section{Design of SAX Navigator}\label{Sec: Design}
SAX Navigator supports the following analysis tasks:

\noindent
\textbf{T1 -- Explore clusters and general data distribution.} Users should be able to explore the cluster space to see general trends and relationships among clusters and get a high-level impression on the data distribution and variability within a cluster (see \autoref{Sec:Global}).

\noindent
\textbf{T2 -- Analyze individual time series within a cluster.} The system needs to support details-on-demand for individual time series to analyze similarities and detect anomalies and errors (see \autoref{Sec:Local}).

\noindent
\textbf{T3 -- Interactive queries based on sketching.}
Users can sketch patterns of interest to find similar data points (see \autoref{sec: query}).

\subsection{Global Patterns}
\label{Sec:Global}

The tree diagram (i.e., dendrogram) of SAX Navigator (\autoref{Sec: tree diagram}) shows the {\it global} structure of the hierarchical clustering result and represents each cluster node as a heat map (\autoref{Sec: Heatmap}), which allows users to identify the general pattern of a cluster (\textbf{T1}).

\subsubsection{Tree Diagram}\label{Sec: tree diagram}
\begin{figure}[t]
  \centering
  \includegraphics[width=0.8\columnwidth]{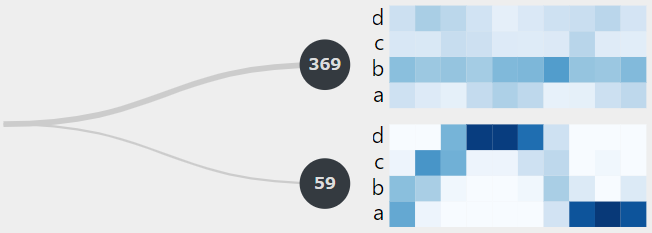}\\
  \textsf{\small (a) Two cluster nodes, their heat maps, and links of the tree diagram.}
  \includegraphics[width=0.8\columnwidth]{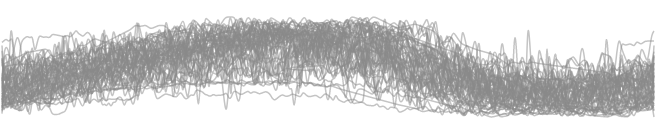}\\
  \textsf{\small (b) Superimposing 59 timelines in the lower cluster of (a).}
  \vspace{-1mm}
  \caption{\footnotesize {Each node in the tree diagram is represented by a circle showing the cluster size and a heat map. A superimposed graph is shown on the cluster detail view.}}
  \vspace{-3mm}
	\label{fig:heat}
\end{figure}
To show the result of the hierarchical clustering, we designed a horizontal node-link tree diagram, as shown in \autoref{fig:teaser} (a). By following links connecting the nodes, users can easily understand how clusters divide into smaller, more similar groups and identify global patterns. 
In the tree diagram, each node is represented by a circle with a number indicating the cluster size and a heat map. 
The cluster size is double-encoded in the link width to the node (see \autoref{fig:heat} (a)).

To address perceptual scalability for a large collection of time series, we filter out small clusters from the tree diagram to reduce visual complexity and clutter.
By default, SAX Navigator shows only clusters whose size is more than 2\% of the total collection. When users want to see more details of a cluster, clicking a node expands the sub-tree of the node. Users are allowed to pan and zoom the tree diagram to explore it at different scales or contexts.

\subsubsection{Heat map for Cluster Aggregation}\label{Sec: Heatmap}
We visually aggregate all time series in a cluster, which are translated into words by SAX, into a heat map display that shows the overall pattern and distribution of the timelines within the cluster without visual clutter (see \autoref{fig:heat} (a)).
In the heat map display, the $x$ axis are the bins ordered by time and the $y$ axis shows the SAX-assigned letter. The color of each cell encodes the proportion of observations with that particular letter assignment at each time slice. The color is on a linear scale that goes from white (no observations) to navy (all observations). 
The lighter the color of a heat map is, as seen in the upper heat map of~\autoref{fig:heat} (a), the more uncertainty or divergence there is in the cluster.
As shown in the lower heat map in~\autoref{fig:heat} (a), the heat map is a much clearer display of the general shape than superimposing all time series in a single line chart like~\autoref{fig:heat} (b).


\subsection{Local Observations}
\label{Sec:Local}


{\it Local} observations are important to understand {\it why} we see certain patterns at a global scale. SAX Navigator supports detailed cluster exploration and local comparisons of $1:1$, $1:n$, and $n:m$ (\textbf{T2}).

\subsubsection{Cluster Detail View}
\label{Sec: Individual display}
%
To analyze individual time series, SAX Navigator can show details-on-demand for all data within a cluster.
Hovering over a cluster node activates the cluster detail view shown in \autoref{fig:teaser} (d). 
The raw time series are shown superimposed on one another in the upper part of the view as well as juxtaposed as sparklines within a data table in the lower part of it. Each row of the table consists of data for a single observation. 
The line chart on top and the rows of the table are connected via brushing and linking. 
Furthermore, by clicking on the rows of the table, SAX Navigator highlights all connecting branches to the observation's ID in the tree diagram.
Users can compare a selected observation in the context of the cluster ($1:n$ comparison) or directly to a second selected observation ($1:1$ comparison).


\subsubsection{Cluster Comparison}
\begin{figure}[t]
    \centering
    \includegraphics[width=.8\columnwidth]{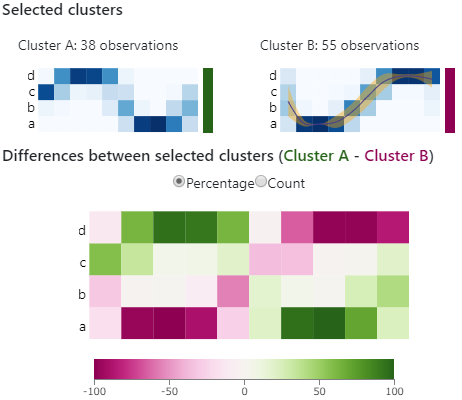}
    \vspace{-2mm}
    \caption{Cluster comparison view. Users can compare two clusters by selecting two of the heat maps within the tree diagram.} 
    \vspace{-3mm}
    \label{fig:heatcompare}
\end{figure}

Users can select two clusters in the tree diagram to start a visual comparison (see \autoref{fig:heatcompare}).
The new heat map shows the differences of the values in the first selected cluster versus the second one. The pattern of Cluster A (left) is colored green, and the pattern of Cluster B (right) is colored magenta. Comparisons can be made in either raw counts or percentages.
Furthermore, we can show the mean and standard deviation of the ``vocabulary of patterns" of the time series in both clusters as a line chart with a confidence interval band, as shown in the upper right heat map of \autoref{fig:heatcompare}. 
The comparison view is particularly helpful for comparing patterns between clusters that are difficult to compare across the tree diagram ($n:m$ comparison).

\subsection{Scalable Query Interface}\label{sec: query}

SAX Navigator provides an interactive sketch-based query interface that allows users to search for observations of interest (\textbf{T3}).

The query tool consists of two options. The first is a drop-down menu where users can select a specific name or ID from the loaded data set. In this case, the path to the selected observation of interest will be highlighted. 
The second method supports searching via user-specified patterns. Inspired by query-by-sketching, we create a grid for users to ``draw" their pattern of interest (i.e., the SAX letter sequence of interest). 
\autoref{fig:teaser} (b) shows a user's selection of an upside down ``V" shape that corresponds to the pattern ``abcba". Using regular expressions, we can quickly search the data set and automatically highlight all tree branches in the tree diagram that contain the specified pattern (see \autoref{fig:teaser} (a)).

\section{Implementation}
SAX Navigator is a web application based on D3.js \cite{d3_framework} and the Flask microframework \cite{flask_framework}. Readers can access a fully interactive prototype at \href{https://sax-navigator.herokuapp.com/}{https://sax-navigator.herokuapp.com/}.

\section{Evaluation}\label{Sec: Usage scenarios}
Our evaluation is based on three case studies and feedback by a domain expert. 
We used 2{,}000 observations from the Catalina surveys data release 2 consisting of 46{,}000 brightness observations~\cite{catalina}, and retrieved commonly used features. The Catalina survey is a well-known and trusted data set about different types of stars. 
Initial feedback from astronomers indicate that they can find search results of interest faster with SAX Navigator than with traditional methods such as table-based feature comparisons.
%


\subsection{Case Studies}

\begin{figure}[t]
    \centering
    \includegraphics[width=.97\columnwidth]{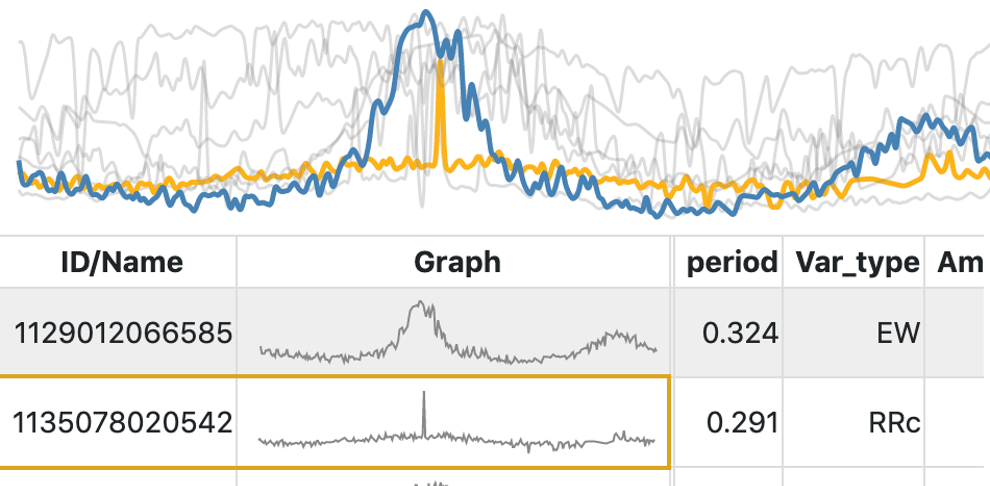}\\
    \textsf{\small (a) $1:1$ comparison.}\\
    \includegraphics[width=.97\columnwidth]{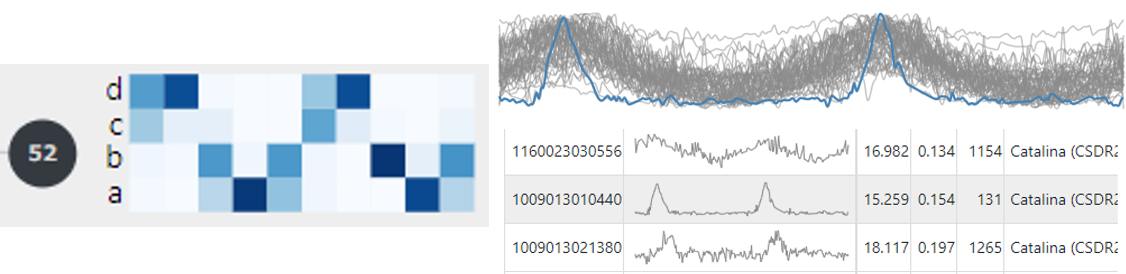}\\
    \textsf{\small (b) $1:n$ comparison.}\\
    \includegraphics[width=.97\columnwidth]{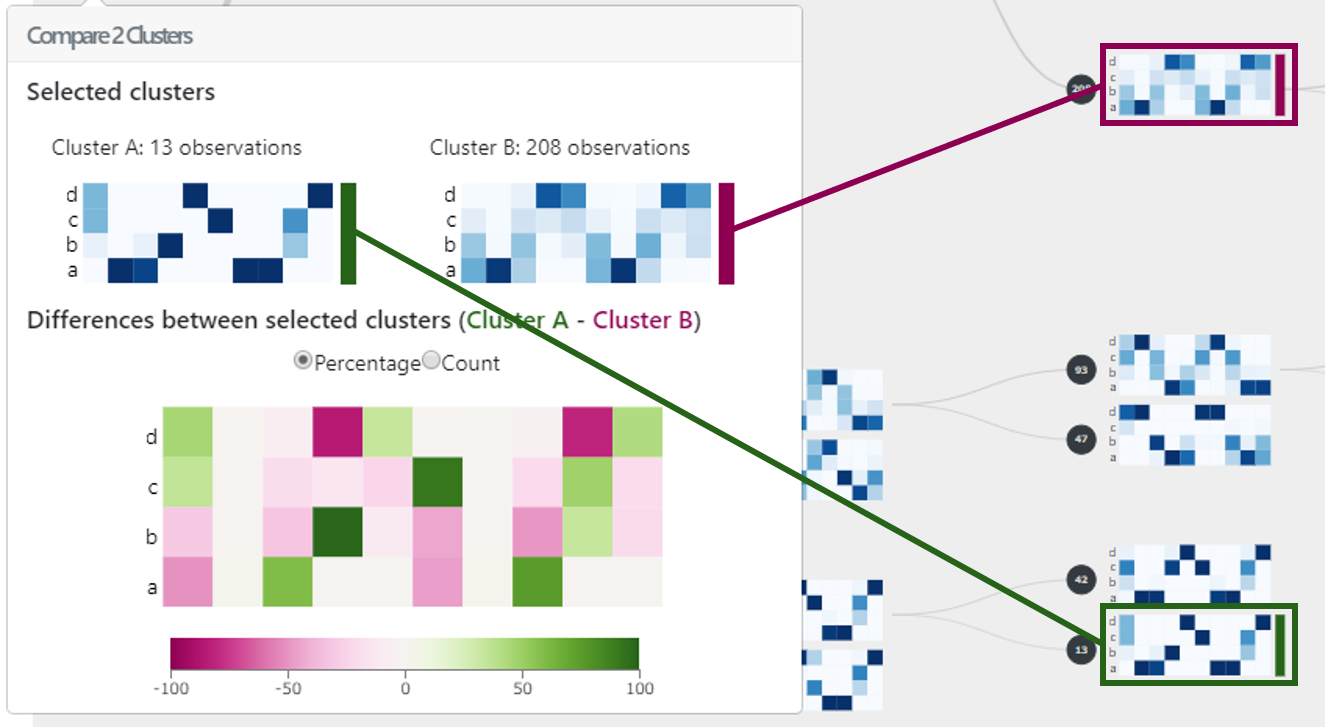}\\
    \textsf{\small (c) $n:m$ comparison.}
     \vspace{-2mm}
    \caption{Case studies of $1:1$, $1:n$, and $n:m$ comparisons. (a) 
    Two interesting observations within a single cluster can be examined and compared in high detail.
    (b) The blue sparkline represents an observation that appears to be incorrectly associated with the cluster. 
    (c) The astronomer can quickly observe the differences between two clusters from completely separate sections of the tree diagram.}
    \vspace{-4mm}
    \label{fig:case studies}
\end{figure}

For astronomical time series clustering, we implemented and used a kernelized cross-correlation distance metric~\cite{Wachman2009KernelsFP} as the primary form of morphological comparison.
Using SAX Navigator, the analyst can discover new patterns and verify the classifications provided for the survey's collection. Let us revisit the example of an astronomer with case studies for our three types of local comparisons. 

\subsubsection{$1:1$ Comparison}
Astronomers frequently compare well-known objects to new observations of interest to classify them. In SAX Navigator, the astronomers can perform a $1:1$ comparison by using the details-on-demand features for local observations in a single cluster. 
For example, to determine whether the gold sparkline seen in \autoref{fig:case studies} (a) is simply noisy or an actual misclassification, users can investigate the data by looking at side-by-side comparisons of the shape of the observations as well as at the metadata of the two selected time series.


\subsubsection{$1:n$ Comparison}
Astronomers have to deal with uncertainties due to instrumental errors related to telescope machinery. These errors can lead to misclassifying the types of celestial observations present in a large astronomy survey. 
For example, suppose an astronomer has identified a cluster in the tree diagram and wants to determine if any of its members have been erroneously assigned due to instrumental error. As seen in \autoref{fig:case studies} (b), the astronomer can hover over the cluster's heat map on the left to view the cluster detail view seen on the right. By hovering over a row in the cluster detail view, the astronomer isolated a data error present in the cluster and highlighted it as a blue sparkline to make a $1:n$ comparison. 
The comparison allowed the astronomer to verify the error's abrupt spikes at the beginning and just after the middle of the timeline when compared to the more gradual increases and decreases seen in the grey sparklines.


\subsubsection{$n:m$ Comparison}

Oftentimes, astronomers explore subtle differences between periodic observations which lead to correct classifications. For example, the heat map comparisons depicted in \autoref{fig:case studies} (c) show the differences at specific points in time between two clusters of periodic observations from separate sections of the hierarchical tree structure. In this instance, the heat map can provide a starting point to understand why one cluster is made up primarily of RR Lyrae variables, while the other additionally contains Cepheids. 
While the heat maps of both clusters show a similar periodic shape, the gaps seen as white and grey space throughout the pattern in Cluster A's heat map suggest that observations were missing at points throughout the timeline. Cluster B shows a fuller pattern which strengthens the astronomer's confidence in the sampling. The larger difference heat map further highlights the points at which Cluster A lacks samples.


\subsection{Domain User Feedback}

To assess the application's usability, we conducted a think-aloud study with an astronomy graduate researcher who has worked with astronomy observation data for 6 years. We gave the participant no suggestions on how to use the system upfront and observed his usage. We answered clarifying questions about the options available and how to pan/zoom on the main visualization. 
The participant first explored the options panel at the top half of the screen. 
The main visualization was most appealing to the participant, he quickly focused his attention on navigating the tree. 
At first, he did not understand how the clusters were formed and suggested that more transparency was needed in the design to explain the distance metric utilized. Once he gained more experience using the tree navigation and had examined specific cluster members in the cluster detail view, his overall response was very positive. 
He stated that \textit{``Wow, this is a great way to quickly see what patterns are in the survey!''} and immediately wanted to load his own data set.
He noted that using the tree diagram and heat map comparison tool enabled him to separate prominent collections of periodic eclipsing binaries. He was able to find subtle differences across these collections at certain points in time, an important and difficult task, much faster when compared to traditional methods like a table-based visualization. 
\section{Conclusion \& Future Work}\label{Sec: Discussion}

We developed an interactive visualization that allows domain experts to explore their time series data in an efficient and meaningful manner. Utilizing the SAX algorithm, we extract a vocabulary of patterns specific to the imported data, which allows for efficient clustering and querying at scale. Our interactive interface gives users the ability to show the overall structure of the hierarchical clustering \emph{and} individual cluster details for thousands of time series.

To generalize our approach to other data and domains, we want to add interactive sliders to change the values for the SAX $\alpha$ and $\omega$ parameters. 
This will allow users to fine-tune the amount of smoothing and clustering.
Furthermore, we want to optimize our implementation in regards to scalability and evaluate how well our visualization scales with up to to millions of observations.


\bibliographystyle{abbrv-doi}

\bibliography{main_bib}
\end{document}